\documentstyle[times,epsfig]{jaa}
\begin{document}
\title[]{Modeling the IDV emissions of the BL Lac Objects with a Langevin type stochastic differential equation}
\author[]%
       {C. S. Leung$^1$\thanks{E-mail:astrosinghk@yahoo.com.hk}, J. Y. Wei $^1$\thanks{Email:wjy@bao.ac.cn}, T. Harko$^2$\thanks{E-mail:harko@hkucc.hku.hk}, Z. Kovacs$^2$\thanks{E-mail:zkovacs@hku.hk} \\
        $^1$National Astronomical Observatories, Chinese Academy of Sciences\\ 20A Datun Road, Chaoyang District, Beijing, China\\
         $^2$Department of Physics and Center for Theoretical and Computational Physics, \\ The University of Hong Kong, Pok Fu Lam Road, Hong Kong, PR China}
\maketitle
\label{firstpage}
\begin{abstract}
In this paper, we introduce a simplified model for explaining the observations of the optical intraday variability (IDV) of the BL Lac Objects. We assume that the source of the IDV are the stochastic oscillations of an accretion disk around a supermassive black hole. The Stochastic Fluctuations on the vertical direction of the accretion disk are described by using a Langevin type equation with a damping term and a random, white noise type force. Furthermore, the preliminary numerical simulation results are presented, which are based on the numerical analysis of the Langevin stochastic differential equation.
\end{abstract}

\section{Introduction}
\label{sec:intro}

Intraday variability is usually defined as the variation of the differential photometric magnitude greater than 3 times of sigma value. However, if the variation is greater than 5 times of sigma value is the best and reliable \nocite{wag95,fan05,alo08} (Wagner and Witzel 1995 ; Fan, J. H. 2005 ; Gupta
\emph{et al.} 2008). There are many observations  for the intraday variation in difference bands, and to explain the observational data many theoretical models have also been proposed for explaining these interesting phenomena. A cellular automaton model was introduced initially by \nocite{kwa98} Kawaguchi et al. (1998), and further developed in  \nocite{tan99} Tang (1999).

It is the purpose of the present paper to propose an alternative model for the explanation of the observed intra-day variability in BL Lac objects and for other similar transient events. The basic physical idea of our model is that the source of the intraday variability can be related to some stochastic oscillations of the disk, triggered by the interaction of the disk with the central supermassive black hole, as well as with a background cosmic environment, which perturbs the disk.  To explain the observed light curve behavior, we develop a model for the stochastic oscillations of the disk, by taking into account the gravitational interaction with the central object, the viscous type damping forces generated in the disk, and a stochastic component which describes the interaction with the cosmic environment.  The mathematical model is formulated in terms of a stochastic, Langevin type differential equation, describing the stochastic oscillation of the accretion disk. This stochastic oscillation model can reproduce the aperiodic light curves associated with transient astronomical phenomena.  Usually the IDV phenomena is explained by using the jet model. There are definitely interactions between the jet and the accretion disk. The stochastic fluctuations of the disk can influence the jet through some energy transfer, and they can represent a source for the jet perturbations.

\section{Vertical thin disk undamped oscillations}

We consider the oscillations of the disk as a whole body under the influence of gravity. The disk is considered thin in the sense discussed by Shakura and Sunyaev (1973). We assume that at the center of the disk we have a compact object of mass $M$. We approximate the distribution of the surface density in the geometrically thin disk by the formulae $\Sigma=\Sigma_{0}={\rm constant},R_{in}\le r\le R_{adj}$, and $\Sigma=\Sigma_0\left(\frac{r}{R_{adj}}  \right)^{-\gamma},R_{adj}\le r\le R_{out}$, respectively, \nocite{ShSu73,ShSu276, Tit}(Shakura and Sunyaev 1973; Shakura  and Sunyaev 1976; Titarchuk and Osherovich 2000), where $R_{in}$ is the innermost radius of the disk, $R_{adj}$ is an adjustment radius in the disk, and $R_{out}$ is the outer radius of the disk. The index $\gamma $ of the surface density can be either $3/5$ or $3/4$.

The vertical oscillations of the disk as a whole can be described by the equation of motion
$M_{d}d^{2}z/dt^{2}+F_{G}(z)=0$, or, equivalently,
\begin{equation}
\frac{d^{2}z}{dt^{2}}+\omega _{0}^{2}z=0,  \label{mot}
\end{equation}
where
\begin{equation}\label{omega}
\omega _{0}^{2}=4\pi ^{2}\nu _{0}^{2}=\frac{\left( 2-\gamma \right) GM}{%
R_{in}R_{adj}^{2}}\left( \frac{R_{out}}{R_{adj}}\right) ^{\gamma -2}\left( 1-%
\frac{\gamma }{\gamma +1}\frac{R_{in}}{R_{adj}}\right) .
\end{equation}

\section{Stochastic fluctuations of accretion disks}

In the following, we consider a model in which we assume that the accretion disk around a massive central object is immersed in a fluctuating fluid flow. For simplicity describe the accretion disk as a macroscopic sphere of radius $a$. In order to describe the stochastic processes in oscillating disks, we
replace the equation of motion of the disk, given by Eq. (\ref{mot}), with a
stochastic Langevin type differential equation, which can be written as
\begin{equation}
\frac{d^{2}z}{dt^{2}}+\xi \frac{dz}{dt}+\omega _{0}^{2}z=\frac{F(t)}{M_{d}},
\label{stoch}
\end{equation}
where the term $\xi dz/dt$, $\xi =$constant, takes into account viscous
dissipation. The fluctuating force $F(t)$ is independent of $z$, and it
varies extremely rapidly as compared to $z$. Since $F(t)$ is very irregular,
we assume that its mean value is zero, $\left\langle F(t)\right\rangle =0$.

By defining
\begin{equation}
E=\frac{M_{d}}{2}\left( \frac{dz}{dt}\right) ^{2}+\frac{1}{2}M_{d}\omega
_{0}^{2}z^{2}-zF(t),
\end{equation}%
as the total energy of the oscillating disk, the luminosity of the disk,
representing the energy lost by the disk due to viscous dissipation and to
the presence of the random force, is given by
\begin{equation}
\frac{dE}{dt}=-L=6\pi \eta H\left( \frac{dz}{dt}\right) ^{2}+z\frac{dF}{dt}.
\end{equation}

As an example of application of this model to a concrete astrophysical system let's consider the case of a thin disk around a compact general relativistic object of mass $M=10^6M_{\odot}$. We assume that the adjustment radius of the disk is $R_{adj}=75R_{in}$, while the outer radius of the disk is located at $R_{out}=250R_{in}$. By assuming $\gamma =3/5$ we obtain first from Eq.~(\ref{omega}) the frequency of oscillations of the unperturbed disk as $\omega _0^2=1.89\times 10^{-8}$ s$^{-2}$. The corresponding oscillation period of the disk in the absence of any dissipative or random processes is $T=18.739$ hours. For the mass of the disk we take  $M_d=1.58\times 10^{34}\;{\rm g}=7.9M_{\odot}$. For the dumping coefficient $\xi $ we obtain $\xi =6\pi H\eta /M_d=2.65\times 10^{-6}$ s$^{-1}$, where for the viscosity coefficient we have assumed the value $\eta =10^{13}$ erg s/cm$^3$. Therefore the equation of motion of the stochastically oscillating disk is given by
\begin{equation}
\frac{d^{2}z}{dt^{2}}+2.65\times 10^{-6} \frac{dz}{dt}+1.89\times 10^{-8}z=6.329\times 10^{-35}F(t).
\label{stoch1}
\end{equation}

The results of the numerical integration of Eq.~(\ref{stoch}) are represented in Figs.~1 and 2.

\begin{figure}[!h]\label{Leung1}
\includegraphics[height=1.798in, width = 2.2in]{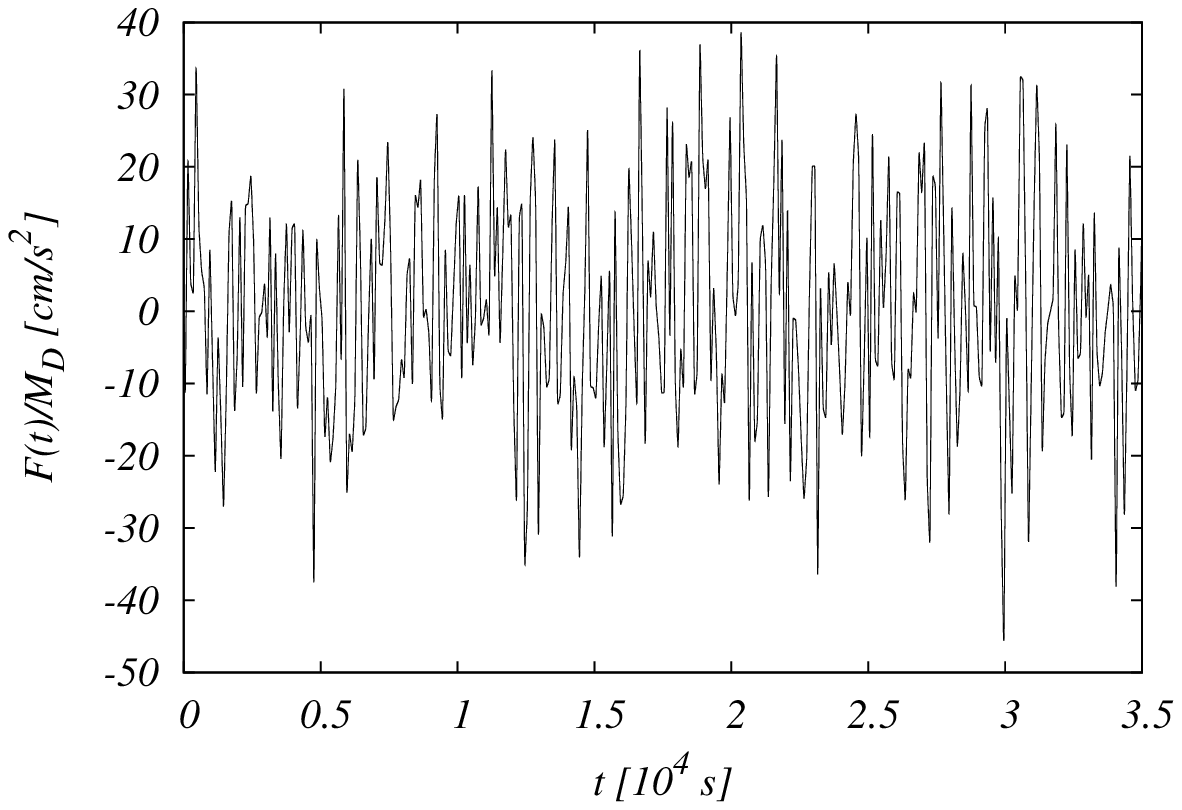}
\includegraphics[height=1.798in, width = 2.2in]{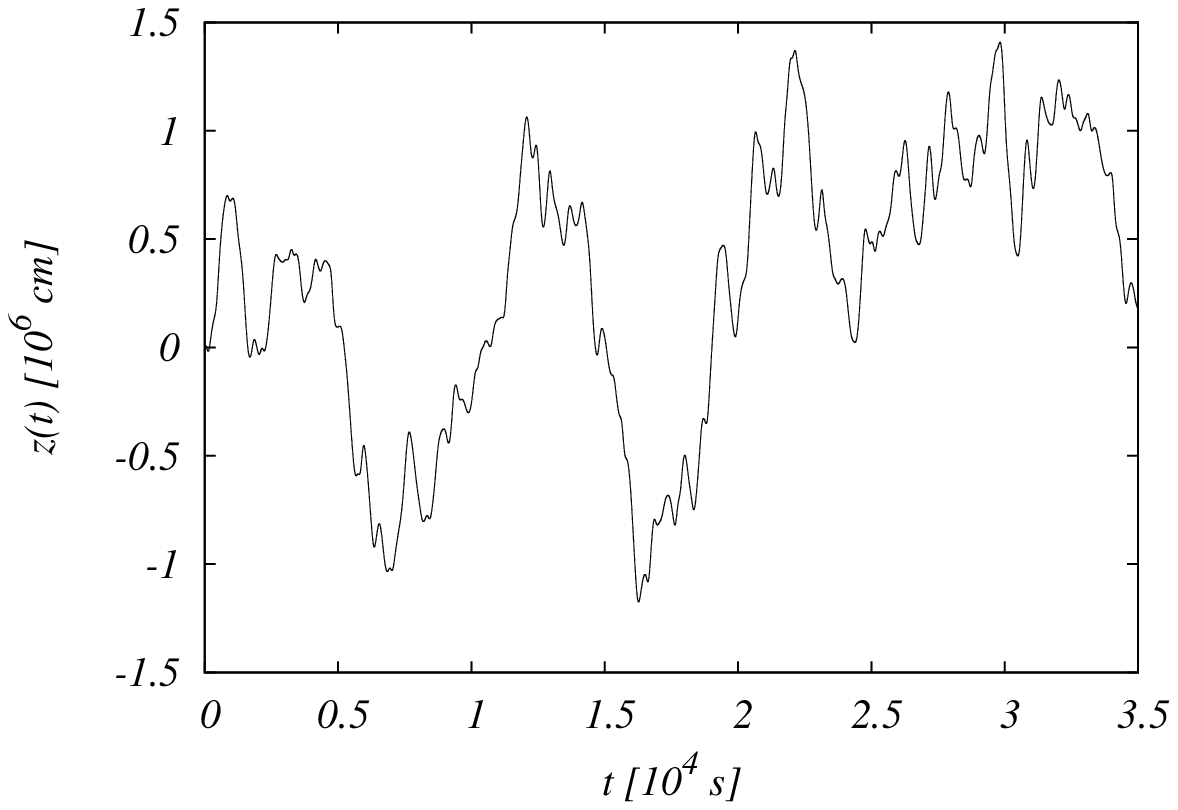}
\caption{The stochastic force $F(t)/M_d$ cm/s$^2$ (left figure) and $z$ (right figure) as a function of time.}
\end{figure}

\begin{figure}[!h]\label{Leung2}
\includegraphics[height=1.798in, width = 2.2in]{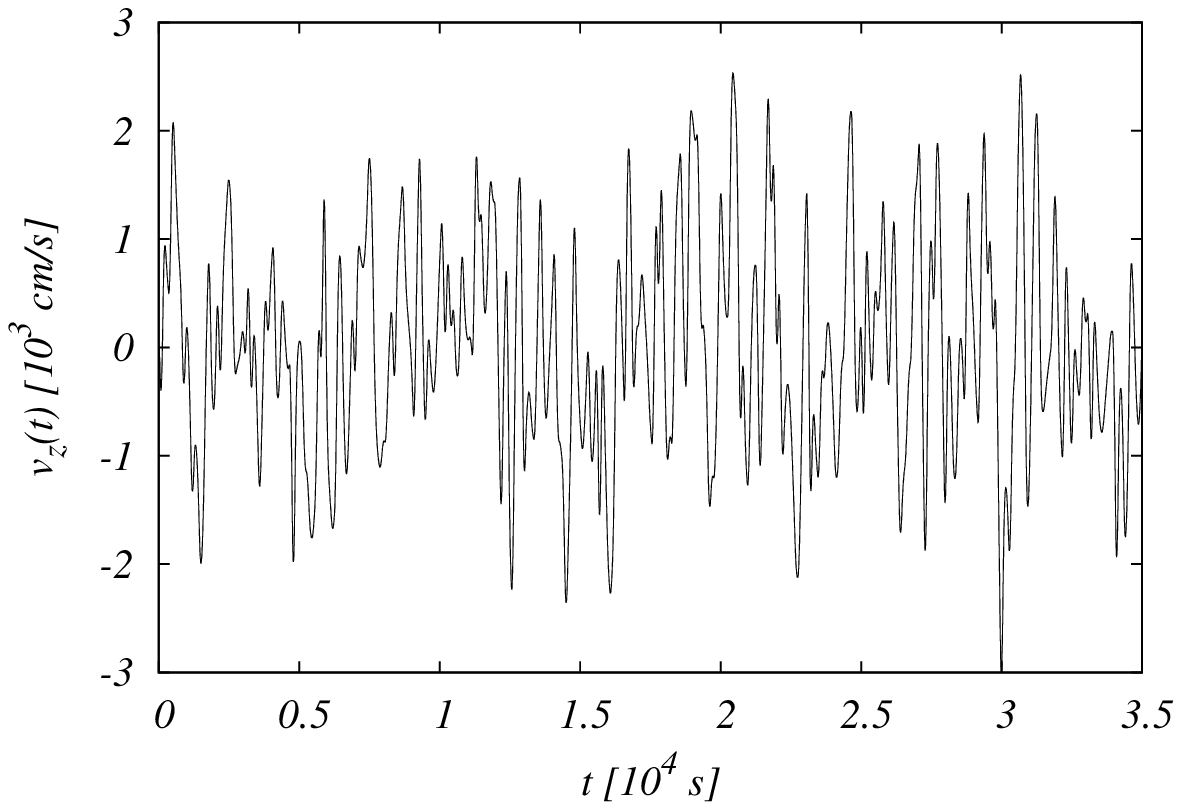}
\includegraphics[height=1.798in, width = 2.2in]{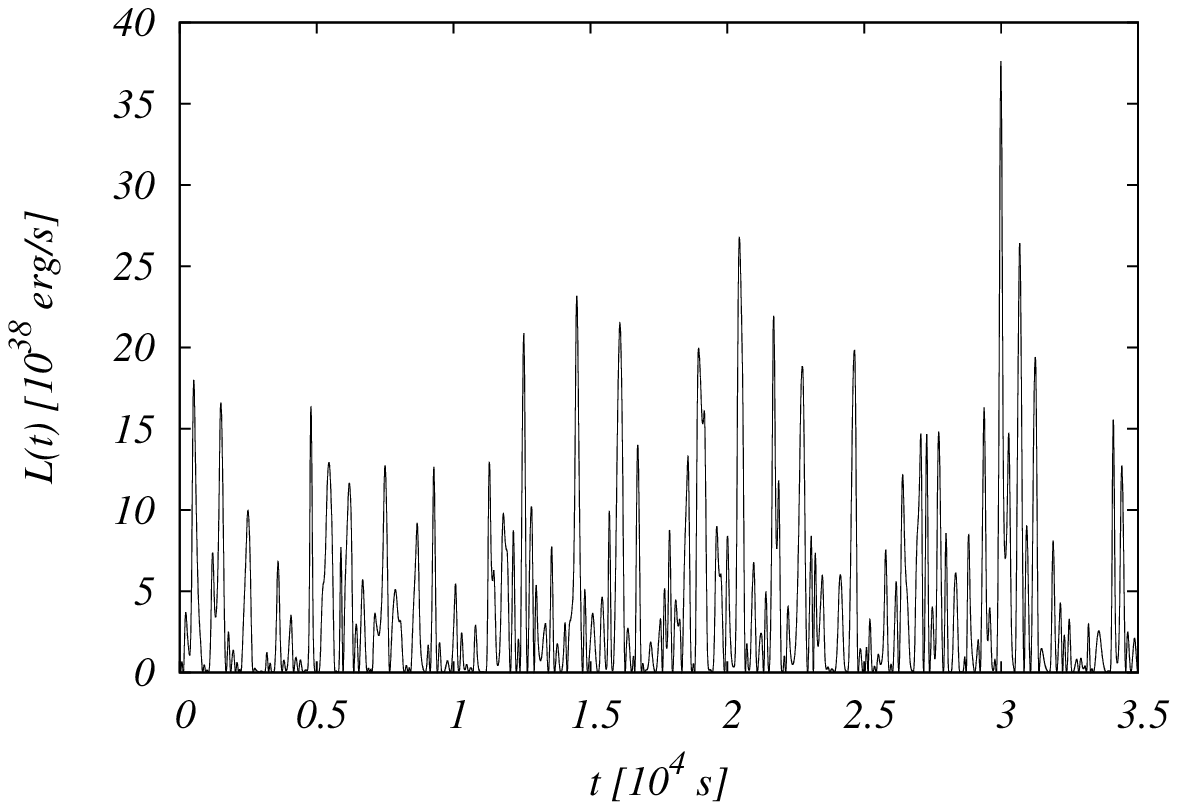}
\caption{The stochastic velocity (left figure) and the luminosity of the disk (right figure) as a function of time.}
\end{figure}

\section{Conclusions}
We have developed a theoretical model describing the stochastic oscillations of the disk. A comparison of the model predictions with the observational data of IDV from different sources will be done in a forthcoming paper. For the case of the ADAF disks, we can also apply the Langevin equation in the vertical direction, since this type of disks also has some vertical oscillation modes. Although the geometry of the ADAF disk is geometrically thick and optically thin, it can also provide a source for the disturbances of the jet.The effect of the jets in the framework of the present model and the effects of the relativistic corrections to the model will be also analyzed.

\end{document}